\documentclass{rjparticle-nolineno}

\usepackage{graphicx}
\usepackage{amsmath}
\usepackage{amssymb}
\usepackage{mathtools}
\usepackage{isotope}
\usepackage{mathrsfs}

\usepackage{liealg}
\usepackage{wignert}
\usepackage{mcmath}



\renewcommand{\topfraction}{.99}
\renewcommand{\bottomfraction}{0.99}
\renewcommand{\textfraction}{0.01}
\newcommand{\Nmax}{{N_\text{max}}}
\newcommand{\scrD}{{\mathscr{D}}}

\newcommand{\scrR}{{\mathscr{R}}}
\newcommand{\MeV}{{\mathrm{MeV}}}
\newcommand{\keV}{{\mathrm{keV}}}

\newcommand{\hw}{{\hbar\Omega}}

\newcommand{\Qvec}{\vec{Q}}

\begin{document}


\title{
Emergence of rotational collectivity in \textit{ab initio} no-core configuration
interaction calculations
}

\author[1]{M. A. Caprio}
\author[2]{P. Maris}
\author[2]{J. P. Vary}
\author[3]{R. Smith}
\affil[1]{Department of Physics, University of Notre Dame,\\ Notre Dame, Indiana 46556-5670, USA}
\affil[2]{Department of Physics and Astronomy, Iowa State University,\\ Ames, Iowa 50011-3160, USA}
\affil[3]{School of Physics and Astronomy, University of Birmingham,\\ Edgbaston, Birmingham, B15 2TT, UK}



\maketitle
\begin{abstract}
Rotational bands have been observed to emerge in \textit{ab initio}
no-core configuration interaction (NCCI) calculations for $p$-shell
nuclei, as evidenced by rotational patterns for excitation energies,
electromagnetic moments, and electromagnetic transitions.  We
investigate the \textit{ab initio} emergence of nuclear rotation in
the \isotope{Be} isotopes, focusing on $\isotope[9]{Be}$ for
illustration, and make use of basis extrapolation methods to obtain
\textit{ab initio} predictions of rotational band parameters for
comparison with experiment. We find robust signatures for rotational
motion, which reproduce both qualitative and quantitative features of
the experimentally observed bands.
\end{abstract}

\maketitle

\section{Introduction}
\label{sec-intro}

Nuclei exhibit a wealth of collective phenomena,
including clustering, rotation, and
pairing~\cite{rowe2010:collective-motion}, which we may now seek to
understand through \textit{ab initio} approaches.  The challenge of \textit{ab initio} nuclear theory is to
quantitatively predict the complex and highly-correlated behavior of
the nuclear many-body system, starting from the underlying
internucleon interactions.  Significant progress has been made in the
$\textit{ab initio}$ description of light nuclei through large-scale
calculations, and signatures of
collective phenomena have now been obtained in \textit{ab initio}
calculations~\cite{wiringa2000:gfmc-a8,dytrych2007:sp-ncsm-dominance,neff2008:clustering-nuclei,shimizu2012:mcsm,dytrych2013:su3ncsm}.
The emergence of rotational bands, in particular, has recently been
observed~\cite{maris2012:mfdn-hites12,maris2012:mfdn-ccp11,caprio2013:berotor,maris2015:berotor2}
in \textit{ab initio} no-core configuration interaction
(NCCI)~\cite{barrett2013:ncsm} calculations for $p$-shell nuclei.
Rotational patterns are found in the calculated level energies,
electromagnetic moments, and electromagnetic transitions.

Yet, NCCI calculations are, of
necessity, carried out in a finite, truncated space.  
Computational restrictions limit the extent to which converged
calculations can be obtained to identify
collective phenomena.  

Therefore, natural questions surrounding the emergence of rotation in
\textit{ab initio} calculations include: \textit{How recognizable is
  the rotation (from the calculated observables)?}  \textit{How robust
  is the rotation (in incompletely converged calculations)?}
\textit{How realistic is the rotation (when quantitatively compared to
  experiment)?}  \textit{How does the rotation arise (or what is the
  intrinsic physical structure)?}  In this proceedings contribution,
we briefly introduce \textit{ab initio} NCCI calculations
(Sec.~\ref{sec-ncci}) and the signatures of nuclear rotation
(Sec.~\ref{sec-rot}).  We then consider the emergence of rotation in
NCCI calculations (Sec.~\ref{sec-rot-be}).  We introduce and augment
the central ideas and discussions of Ref.~\cite{maris2015:berotor2},
touching on the above questions, using $\isotope[9]{Be}$ for
illustration.  We then compare the rotational energy parameters of the
calculated bands in $\isotope[7\text{--}12]{Be}$ against those of
their experimental counterparts.  We apply an exponential basis
extrapolation scheme for the energies, to estimate the band parameters
which would be found in an untruncated calculation.

\section{NCCI calculations}
\label{sec-ncci}

In NCCI calculations, the nuclear many-body Schr\"odinger equation 
is formulated as a Hamiltonian matrix eigenproblem.  The Hamiltonian
is represented with respect to a basis of antisymmetrized products of
single-particle states, conventionally, harmonic oscillator states.
The problem is then solved for the full system of $A$ nucleons,
\textit{i.e.}, with no inert core.  However, calculations must
be carried out in a finite-dimensional space, commonly obtained by
truncating the basis to a maximum allowed number $\Nmax$ of oscillator
excitations.  Convergence toward the exact results~--- as would be
achieved in the full, infinite-dimensional space~--- is obtained with increasing $\Nmax$.  However, the basis
size grows combinatorially with $\Nmax$, so the maximum accessible
$\Nmax$ is severely limited by computational restrictions.

\begin{figure}[t]
\centerline{\includegraphics[width={0.95}\hsize]{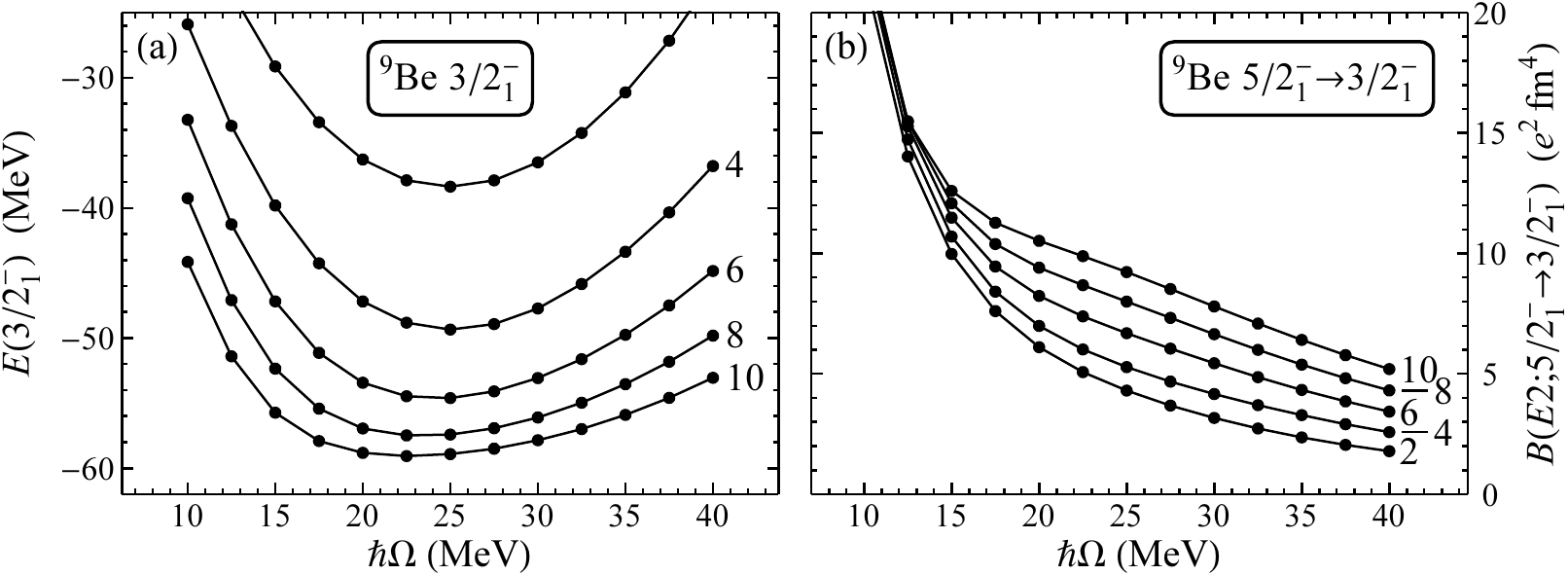}}
\caption{The
  $\Nmax$ and $\hw$ dependence of calculated observables for
  $\isotope[9]{Be}$: (a)~the energy of the $3/2^-$ ground state and
  (b)~the yrast $5/2^-\rightarrow 3/2^-$ electric quadrupole
  transition strength.
  Calculated values are shown as functions of $\hw$ for
  $\Nmax=2$ to $10$ (as
  labeled).}
\label{fig-9be-hw}      
\end{figure}

The calculated eigenvalues and wave functions, and thus observables,
depend both upon the basis truncation $\Nmax$ and on the length
parameter for the basis, which is specified by the oscillator energy
$\hw$.  These dependences are illustrated for the calculated
ground-state energy of $\isotope[9]{Be}$ in Fig.~\ref{fig-9be-hw}(a),
for $2\leq \Nmax\leq10$.  For a fixed $\Nmax$, a minimum
in the calculated energy is obtained in the range $\hw\approx20\,\MeV$
to $25\,\MeV$, providing a variational upper bound.  Then, as $\Nmax$ is increased, a lower
calculated ground state energy is obtained at each $\hw$.  The approach to
convergence is marked by approximate $\Nmax$ independence (a
compression of successive energy curves) and $\hw$ independence (a
flattening of each curve around its minimum).  While a high level of
convergence (at the $\keV$ scale) may be obtained in the lightest
nuclei (in particular, see Fig.~3 of Ref.~\cite{caprio2014:cshalo} for
$\isotope[4]{He}$), the situation is challenging for the
$\isotope{Be}$ isotopes.  Although the decrease in the variational
minimum energy for $\isotope[9]{Be}$ becomes smaller with each step in
$\Nmax$ [Fig.~\ref{fig-9be-hw}(a)], even at $\Nmax=10$ these changes
are still at the $\MeV$ scale.

For electric quadrupole transition strengths, convergence is even more
elusive, as shown for the transition between the lowest $5/2^-$ state and
$3/2^-$ ground state of $\isotope[9]{Be}$ in
Fig.~\ref{fig-9be-hw}(b).
Convergence~--- as manifested in $\Nmax$ independence and $\hw$
independence~--- would here be reflected in a compression of
successive $\Nmax$ curves and a ``shoulder'' in the plot of
the $B(E2)$ against $\hw$ (see Fig.~3 of
Ref.~\cite{caprio2014:cshalo} for analogous convergence, albeit of a different observable, in
$\isotope[4]{He}$).  However, at most hints of  such features are
apparent at $\Nmax=10$ in Fig.~\ref{fig-9be-hw}(b).  Consequently, there is no obvious way to extract a value for the $E2$
strength (other than to arbitrarily
choose an $\hw$ value at which to do so, as we shall consider in Sec.~\ref{sec-rot-be}).

\section{Collective nuclear rotation}
\label{sec-rot}

Nuclear rotation~\cite{rowe2010:collective-motion} arises when there is an adiabatic separation of a rotational degree of
freedom from the remaining internal degrees of freedom of the nucleus.  A rotational state factorizes into an
\textit{intrinsic state} $\tket{\phi_K}$ and a rotational
wave function of the Euler angles $\vartheta$, describing the
collective rotational motion of this intrinsic state.  (Specifically, we
consider an axially symmetric 
intrinsic state, with definite angular momentum projection $K$ along
the intrinsic symmetry axis.)  The full nuclear state $\tket{\psi_{JKM}}$,
with total angular momentum $J$ and projection $M$, has the form
\begin{equation}
\label{eqn-psi}
\tket{\psi_{JKM}}\propto
\int d\vartheta\,\bigl[
~
\underbrace{\scrD^J_{MK}(\vartheta)}_{\text{Rotational}}
~
\underbrace{\tket{\phi_K;\vartheta}}_{\text{Intrinsic}}
~
+
~
(-)^{J+K}\scrD^J_{M-K}(\vartheta)\tket{\phi_{\bar{K}};\vartheta}
\bigr],
\end{equation}
where $\tket{\phi_K;\vartheta}$ represents the intrinsic state
$\tket{\phi_K}$ after
rotation by $\vartheta$, and the second term involving the $\scrR_2$-conjugate state
$\tket{\phi_{\bar{K}};\vartheta}$ arises from discrete rotational symmetry considerations.  

The principal signatures of rotational structure reside in rotational
patterns in the energies and electromagnetic multipole observables for
the set of rotational states.  Rotational band members, sharing the
same intrinsic state but differing in $J$ and hence their rotational
wave functions, have energies following the rotational formula
$E(J)=E_0+AJ(J+1)$, where the rotational energy constant $A$ is
inversely related to the moment of inertia of the intrinsic state.  However, for $K=1/2$
bands, the Coriolis contribution to the kinetic energy significantly
modifies this pattern, yielding an energy staggering
\begin{equation}
\label{eqn-EJ-stagger}
E(J)=E_0+A\bigl[J(J+1)+
\underbrace{a(-)^{J+1/2}(J+\tfrac12)}_{\text{Coriolis ($K=1/2$)}}
\bigr],
\end{equation}
where $a$ is the Coriolis decoupling parameter.
Electromagnetic transitions among the bandmembers
likewise follow a well-defined pattern.  For the electric quadrupole operator, in particular, the reduced matrix element~\cite{edmonds1960:am}
between initial and final band members with angular momenta  $J_i$ and $J_f$, respectively, follows
the relation
\begin{equation}
\label{eqn-MEE2}
\trme{\Psi_{J_fK}}{\Qvec_2}{\Psi_{J_iK}}
= 
(2J_i+1)^{1/2}
~
\underbrace{\tcg{J_i}{K}{2}{0}{J_f}{K}}_{\text{Rotational}}
~
\underbrace{\tme{\phi_K}{Q_{2,0}}{\phi_K}}
_{\text{Intrinsic ($\propto eQ_0$)}}.
\end{equation}
The value depends on the particular band members involved only
through a Clebsch-Gordan coefficient, while the details of the
intrinsic state determine  the intrinsic quadrupole moment
$eQ_0\equiv(16\pi/5)^{1/2}\tme{\phi_K}{Q_{2,0}}{\phi_K}$.  Quadrupole
moments $Q(J)$, proportional to
$\trme{\psi_{JK}}{\Qvec_2}{\psi_{JK}}$, and reduced transition
probabilities $B(E2;J_i\rightarrow J_f)$, proportional to
$\abs{\trme{J_f}{\Qvec_2}{J_i}}^2$, are uniquely related by~(\ref{eqn-MEE2}), to
within a common normalization determined by $Q_0$.  

\section{Rotation in the \NoCaseChange{Be} isotopes}
\label{sec-rot-be}

\begin{figure}[tp]
\centerline{\includegraphics[width={0.95}\hsize]{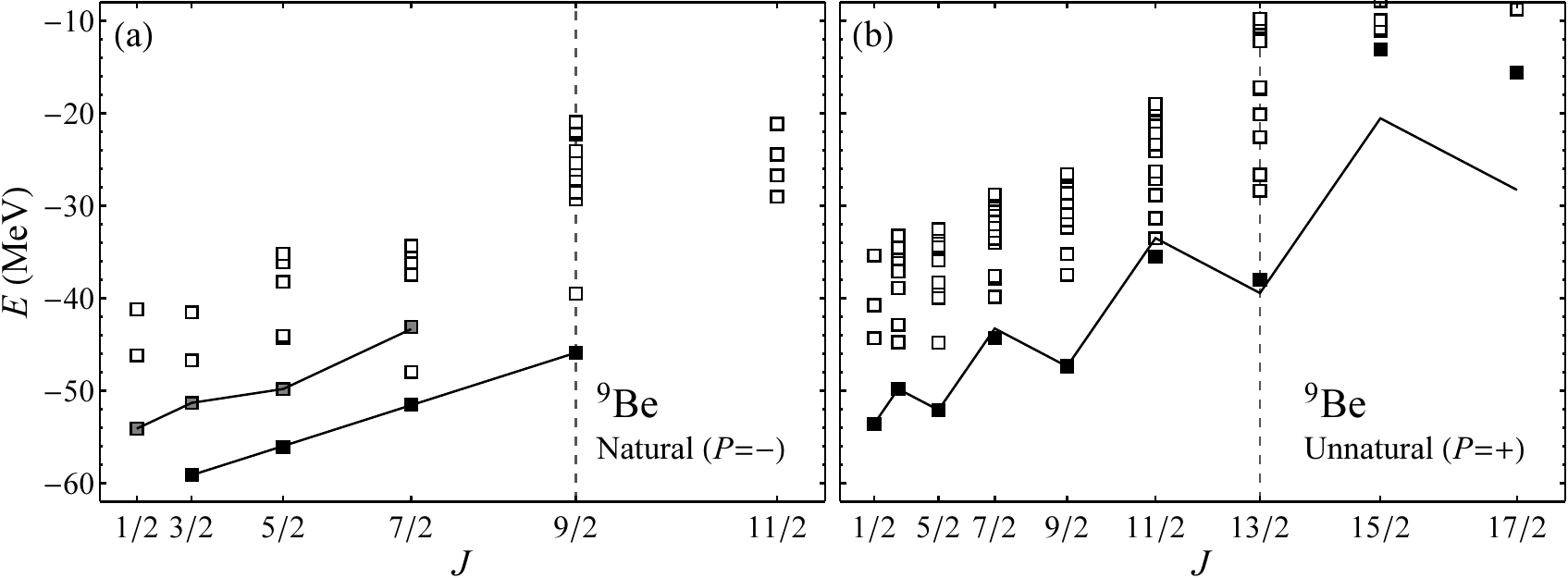}}
\caption{Energy eigenvalues obtained for states in  the
natural~(left) and unnatural~(right) parity spaces for
$\isotope[9]{Be}$.  Energies are plotted with
respect to an angular momentum axis which is scaled to be linear in
$J(J+1)$.
Solid symbols indicate candidate 
yrast band members, and shaded symbols indicate candidate excited band
members.  Solid lines
indicate the corresponding fits for rotational energies~(\ref{eqn-EJ-stagger}) (based on the
first three band members for $K=1/2$ or all band members for $K=3/2$).  
Vertical dashed lines indicate the maximal angular
momentum accessible within the lowest harmonic oscillator
configuration (or valence space).}
\label{fig-energy-9be}      
\end{figure}

Energies following a rotational pattern are most easily recognized if
plotted against an angular momentum axis which is scaled as $J(J+1)$,
so that energies in an ideal rotational band lie on a straight line
(or staggered about a straight line, for $K=1/2$).  Such a plot of
energies against angular momentum is shown for the calculated states
of $\isotope[9]{Be}$ in Fig.~\ref{fig-energy-9be}, both for natural
(negative) parity states~(left) and unnatural (positive) parity
states~(right). (The parity of the lowest allowed oscillator
configuration, or traditional shell model valence space, may be termed
the \textit{natural parity}, and that obtained by promoting one
nucleon by one shell the \textit{unnatural parity}.)

These calculations, from Ref.~\cite{maris2015:berotor2}, are obtained
taking a specific basis truncation $\Nmax=10$ (for natural parity, or
$\Nmax=11$ for unnatural parity) and basis parameter $\hw=22.5\,\MeV$
(near the variational minimum). They thus may be thought of as taking
a snapshot of the spectrum along the path to convergence.  These
calculations are based on the realistic JISP16 nucleon-nucleon
interaction~\cite{shirokov2007:nn-jisp16}.  The Coulomb interaction
has been omitted, to ensure exact conservation of isospin, thereby
simplifying the spectrum (we consider states of minimal isospin
$T=T_z$ only).  However, the effect of the Coulomb interaction, if
included, is mainly to induce an overall shift in the binding
energies.

Rotational bands are most readily identifiable near
the yrast line, where the density of states remains comparatively low.
Identification of band members is based not only on recognizing
rotational energy patterns, but also on observing collective
enhancement of electric quadrupole transition strengths among band
members and verifying rotational patterns of electromagnetic moments
and transitions.  In the natural parity space [Fig.~\ref{fig-energy-9be}(a)], a $K=3/2$
yrast band (solid symbols) and $K=1/2$ excited band (shaded symbols)
are identified, while, in the unnatural parity space
[Fig.~\ref{fig-energy-9be}(b)], a $K=1/2$ yrast band is identified
(solid symbols).  For each of the candidate bands, the line
indicates a rotational energy fit to the calculated band members.  

Intriguing features of the bands found in the $\isotope{Be}$ isotopes
relate to their termination at finite angular momentum.  Some bands
terminate at (or below) the maximal angular momentum permitted within
the valence space, as in the case of both natural parity bands of
$\isotope[9]{Be}$ [Fig.~\ref{fig-energy-9be}(a)].  [The maximal
  angular momentum permitted within the valence space (or lowest
  harmonic oscillator configuration) of each parity is indicated by
  the vertical dashed lines in Fig.~\ref{fig-energy-9be}.]  However,
some bands also extend beyond this angular momentum, as in the case of
the unnatural parity yrast band of $\isotope[9]{Be}$
[Fig.~\ref{fig-energy-9be}(b)].  While this band deviates from the
rotational energy formula above the maximal valence angular momentum,
other calculated bands in the $\isotope{Be}$ isotopes continue through
this angular momentum with no apparent deviation (see
Ref.~\cite{maris2015:berotor2} for examples).

\begin{figure}[t]
\centerline{\includegraphics[width=0.95\hsize]{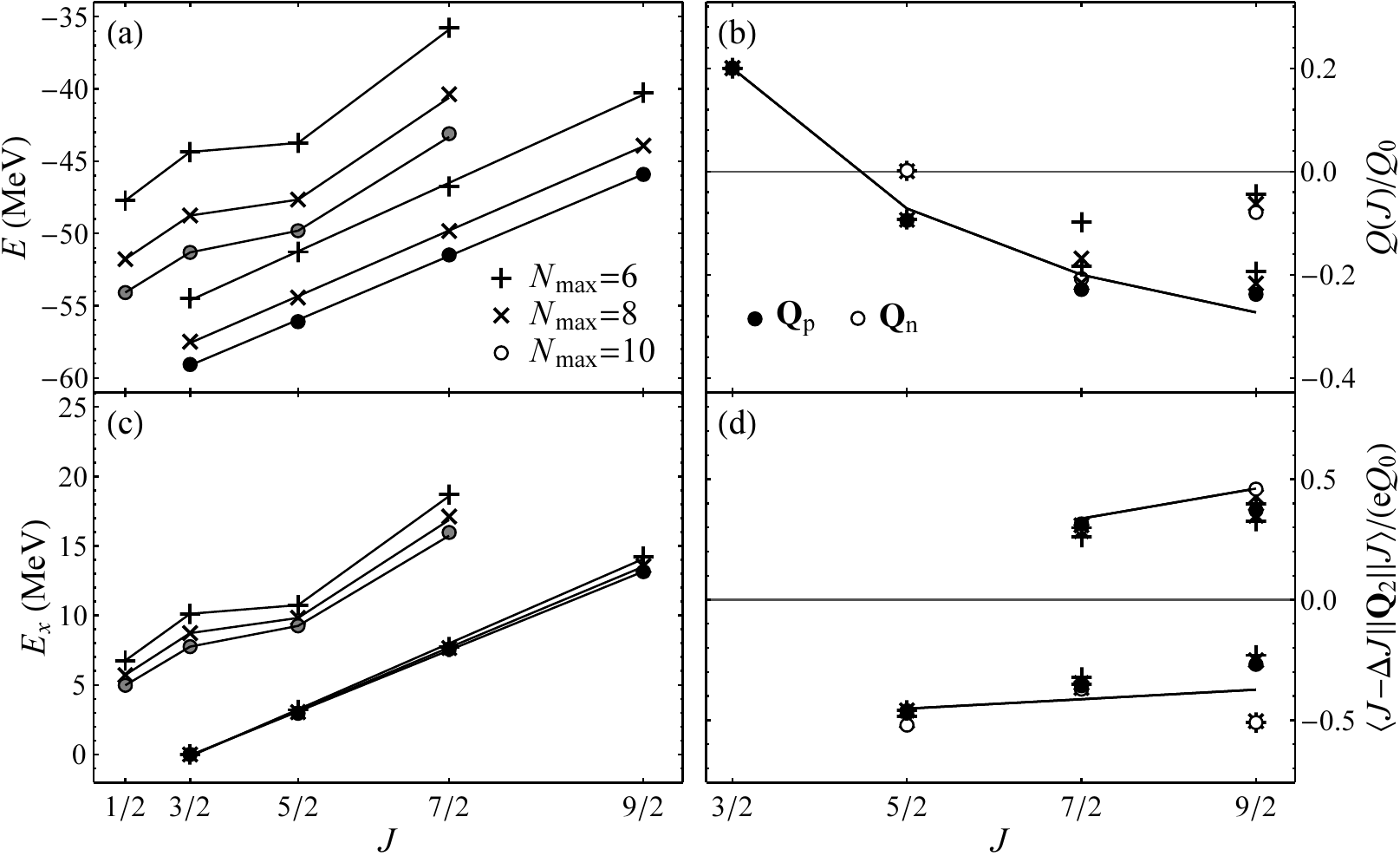}}
\caption{Dependence of calculated observables on the basis truncation
  $\Nmax$, for rotational band members in $\isotope[9]{Be}$:
  yrast and excited band energies, as
  (a)~absolute energies $E$ and (c)~excitation energies $E_x$,  and yrast band
  electric quadrupole observables, both (b)~quadrupole moments and (d)~in-band transition matrix elements.
Values are shown for successive $\Nmax$ truncations ($6\leq \Nmax \leq 10$) to
provide an indication of convergence (calculations of increasing
$\Nmax$ are indicated by plusses, crosses, and circles,
respectively).
Electric quadrupole moments and matrix
  elements are normalized to the intrinsic quadrupole moment $Q_0$, as
  determined from $Q(3/2)$.  The curves indicate the values
  expected from the rotational
  formula.  In panel~(d), the upper and lower curves are for $\Delta J=2$ and $\Delta
J=1$ transitions, respectively (for the definition and significance of the signs of these matrix
elements, see
Ref.~\cite{maris2015:berotor2}).    
}
\label{fig-9be0-conv}      
\end{figure}

\begin{figure}[tp]
\centerline{\includegraphics[width={0.95}\hsize]{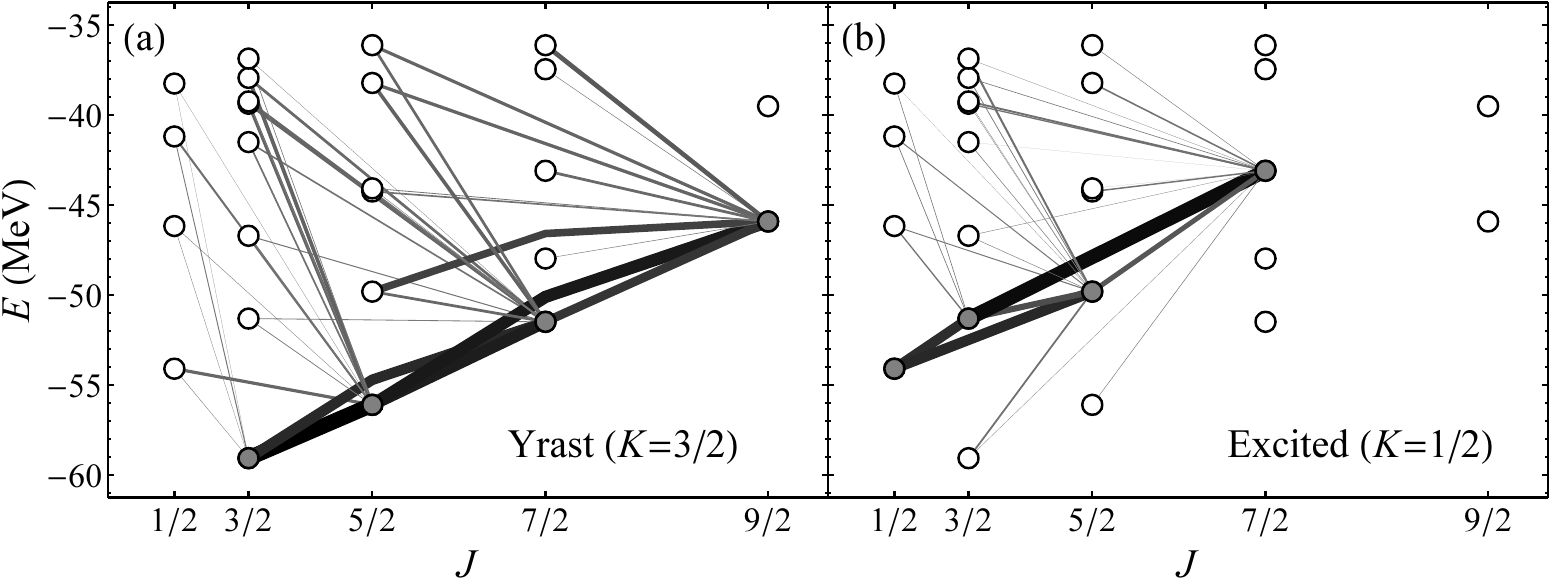}}
\caption{Calculated electric quadrupole transition matrix elements among
  states in the $\isotope[9]{Be}$ natural parity space: (a)~from yrast
  band members and (b)~from excited band members.
  In each panel, shaded circles indicate the initial levels being
  considered for transitions.  All angular momentum
  \textit{decreasing} transitions from the selected levels are shown.
  Line thicknesses are proportional to the magnitude of the 
  reduced matrix element for the transition (also conveyed through a
  gray scale).  Matrix elements are shown for the
  proton quadrupole tensor, but results with the neutron
  quadrupole tensor are similar.
}
\label{fig-trans-9be0-network}      
\end{figure}

Let us return to the challenge of convergence, now in relation to
rotational patterns.  The $\Nmax$ dependence of the energy
eigenvalues, at fixed $\hw=22.5\,\MeV$, is shown for the members of
both natural parity bands in Fig.~\ref{fig-9be0-conv}(a).  For each
step in $\Nmax$, the calculated energies shift lower by several
$\MeV$.  However, it may also be seen that the energies of different
members of the same band move downward in approximate synchrony as
$\Nmax$ increases. Thus, the energies within the band remain
comparatively unchanged, as is seen more directly when we consider excitation
energies in Fig.~\ref{fig-9be0-conv}(c).  The excitation energy of the
excited band relative to the yrast band, though not converged, varies
much less rapidly with $\Nmax$ than do the eigenvalues themselves.

The strengths of the various electric quadrupole transitions originating
from the candidate band members in
the natural parity bands of $\isotope[9]{Be}$ are shown in
Fig.~\ref{fig-trans-9be0-network}.  We observe both the dominance of
transitions within each band and the nonnegligible cross-talk between
bands, as well as to some states outside these bands.

Although the calculated electric quadrupole observables are far from
converged [Fig.~\ref{fig-9be-hw}(b)], rotational structure \textit{per
  se} is reflected in the \textit{ratios} of matrix elements within a
band, not their absolute magnitudes, which depend also upon the intrinsic structure via $Q_0$ 
[see~(\ref{eqn-MEE2})].   In Fig.~\ref{fig-9be0-conv} (at right), we consider the calculated electric quadrupole matrix elements
within the $\isotope[9]{Be}$ yrast band~--- quadrupole moments [Fig.~\ref{fig-9be0-conv}(b)]
and quadrupole transition matrix elements (both $\Delta J=1$ and $\Delta J=2$) [Fig.~\ref{fig-9be0-conv}(d)]~--- and compare with the rotational
predictions (lines).  The overall normalization $Q_0$ is
eliminated, by normalizing to 
the quadrupole moment of the band head [from which a value of $Q_0$ is
  determined by~(\ref{eqn-MEE2})].   Thus, rather than considering
the unconverged matrix elements individually, as in Fig.~\ref{fig-9be-hw}(b), we effectively only consider
\textit{ratios}, of one unconverged matrix element
to another.
These normalized matrix
elements are seen to be comparatively independent of $\Nmax$, at the $\Nmax$ values
considered.  

The resemblance between the calculated quadrupole moments and
transition matrix elements and the expected rotational values
(Fig.~\ref{fig-9be-hw}(b,d)), while clearly not perfect, is strongly
suggestive of rotation. One should bear in mind that quadrupole
moments of \textit{arbitrarily} chosen states in the spectrum
fluctuate not only in magnitude but also in sign, and that calculated
transition matrix elements for arbitrarily chosen pairs of states
fluctuate by many orders of magnitude
[Fig.~\ref{fig-trans-9be0-network}].  The rotational
relation~(\ref{eqn-MEE2}) is equally valid whether we take the
quadrupole operator to be the proton quadrupole tensor (\textit{i.e.},
the physical electric quadrupole operator) $\Qvec_p$ or the neutron
quadrupole tensor $\Qvec_n$.  The matrix elements of these two
operators, shown together in Fig.~\ref{fig-9be0-conv}, provide
valuable complementary information for investigating rotation.  More
extensive examples, including bands which extend to higher angular
momentum, and considering magnetic dipole observables, may be found in
Ref.~\cite{maris2015:berotor2}.

\begin{figure}[t]
\centerline{\includegraphics[width={1.0}\hsize]{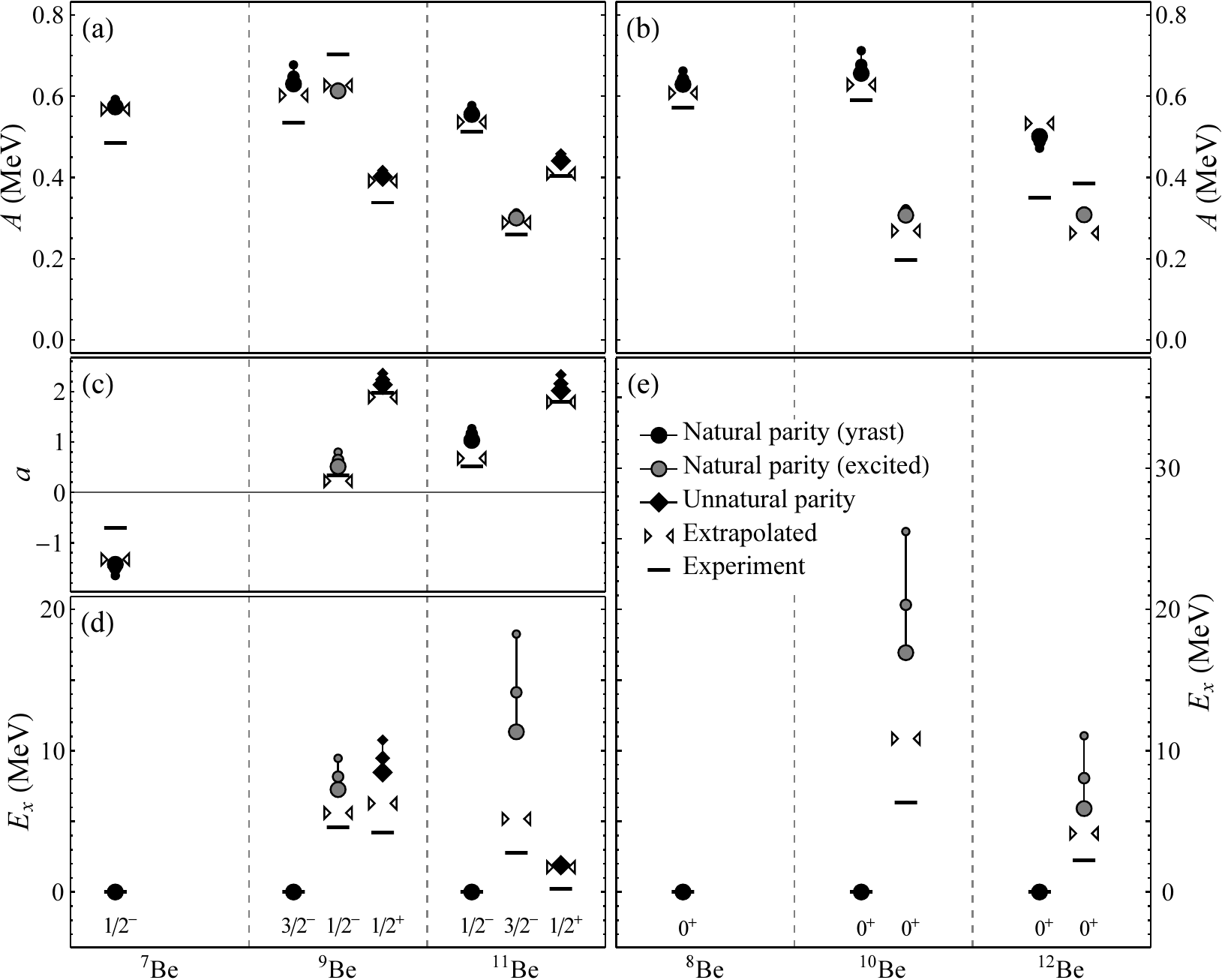}}
\caption{Band energy parameters for odd-mass $\isotope{Be}$
  isotopes~(left) and even-mass $\isotope{Be}$ isotopes~(right): the rotational constant $A$~(top),
  Coriolis decoupling parameter $a$~(middle), and band
  excitation energy $E_x$~(bottom).  Bands are distinguished as
  natural parity yrast (circles, solid), natural parity excited
  (circles, shaded), and unnatural parity yrast (diamonds).  Values
  are shown for $6\leq \Nmax \leq 10$ for natural parity or $7\leq
  \Nmax \leq 11$ for unnatural parity, with larger symbols for higher
  $\Nmax$ values.  Parameter values are also shown based on
  exponentially extrapolated level energies (paired triangles) and experimental
  bands (horizontal lines).  }
\label{fig-params}      
\end{figure}

Returning to the initial questions (from Sec.~\ref{sec-intro}), now
that we have explored how \textit{recognizable} the signatures of
rotation are, let us quantitatively examine how \textit{robust} the
rotational predictions are and how \textit{realistic} they are in
comparison to experiment.  The energy parameters for bands across the
$\isotope{Be}$ isotopic chain are summarized in Fig.~\ref{fig-params}:
the band excitation energy $E_x$ (defined relative to the yrast band
as $E_x\equiv E_0-E_{0,\text{yrast}}$), the band
rotational parameter or slope $A$, and the band Coriolis decoupling
parameter or staggering $a$ (for $K=1/2$).  Results are shown for a
sequence of $\Nmax$ truncations.  Parameters for experimentally
observed bands~\cite{bohlen2008:be-band} are also shown (horizontal
lines).  (The calculations and experimental data are detailed in
Ref.~\cite{maris2015:berotor2}.)

Our results for the rotational parameter $A$ and Coriolis decoupling
coefficient $a$ [Fig.~\ref{fig-params}(a--c)] are sufficiently
stable with respect to $\Nmax$ to permit at least a rough comparison with experiment.  The value of the
rotational parameter [Fig.~\ref{fig-params}(a,b)] varies by a factor of $\sim2$
across the different calculated bands.  The calculations reproduce both
the general range of experimental values for $A$ and the general trend of which
bands have higher or lower slopes (excepting $\isotope[12]{Be}$,
where the discrepancies may arise due to mixing of the yrast and excited bands~\cite{fortune1994:12be-tp}).
The Coriolis staggering for the calculated $K=1/2$ bands [Fig.~\ref{fig-params}(c)] varies 
in both amplitude and sign, and the experimental trend in both these properties is
reproduced across the bands.  

However, the excitation energies $E_x$ [Fig.~\ref{fig-params}(d,e)]
converge at rates which vary considerably among the bands.  In
general, the excitation energies are decreasing with $\Nmax$, bringing
them toward the experimental values.  Yet, they are varying too
strongly with $\Nmax$ for it to be immediately obvious how close the
converged predictions will lie to the experimental values.

More detailed comparisons with the experimentally identified
rotational bands could be made possible by the application of basis extrapolation
methods~\cite{maris2009:ncfc,coon2012:nscm-ho-regulator,furnstahl2012:ho-extrapolation},
which  are being developed to deduce the
converged values for energies (and other observables).  Such
methods are still in their formative stages.  Nonetheless, it is
intriguing to apply a straightforward scheme based on a presumed exponential convergence of energy eigenvalues
with $\Nmax$~\cite{maris2009:ncfc}.  The calculated energies
are taken to approach the converged value $E_\infty$, for $\Nmax\rightarrow\infty$,  as
\begin{equation}
\label{eqn-exp}
E(\Nmax,\hw)=E_\infty+a_{\hw}\exp(-b_{\hw}\Nmax).
\end{equation}
Calculations of the energy at three successive $\Nmax$ values, for fixed $\hw$, are sufficient to determine all three
parameters ($a_{\hw}$, $b_{\hw}$, and $E_\infty$) in~(\ref{eqn-exp}) and
thus provide an extrapolation to the converged energy.  

Using 
extrapolated energies for the band members, we again determine the band energy parameters
(paired triangles in Fig.~\ref{fig-params}), yielding results which
match experiment to a greater (\textit{e.g.}, $\isotope[9]{Be}$
excitation energies) or lesser (\textit{e.g.}, $\isotope[10]{Be}$
excitation energy) degree.    However, such extrapolations are subject to considerable
uncertainties~\cite{maris2009:ncfc}.  It is therefore not yet clear to
what extent the remaining discrepancies reflect
deficiencies in the \textit{ab initio} description of the nucleus, with
the chosen
interaction (JISP16), or limitations of
the exponential extrapolation.

In conclusion, we have seen that, despite the challenges of obtaining
converged energies or electromagnetic observables, rotational patterns
can robustly emerge in \textit{ab initio} NCCI calculations, and these
calculations can yield predictions of rotational properties for both qualitative and
quantitative comparison with experiment.
What remains to be addressed, from among our initial questions (in
Sec.~\ref{sec-intro}), is the physical origin and
\textit{intrinsic structure} of the rotation: prospective
contributors include multishell $\grpsu{3}$ symmetry (shown to be dominant in NCCI calculations~\cite{dytrych2013:su3ncsm}), symplectic
symmetry~\cite{rowe1985:micro-collective-sp6r,dytrych2007:sp-ncsm-dominance}, and 
$\alpha$-$\alpha$ molecular rotation~\cite{kanadaenyo1999:10be-amd}.

\begin{acknowledgement}
Discussions with M.~Freer are gratefully acknowledged.  This work was
supported by the US DOE under Grants
No.~DE-FG02-95ER-40934, DESC0008485 (SciDAC/NUCLEI), and
DE-FG02-87ER40371, by the US NSF under
Grant No.~0904782, and by the Research Corporation for Science
Advancement Cottrell Scholar program. 
Computational resources were provided by the
National Energy Research Scientific Computing Center (NERSC)
and the Argonne
Leadership Computing Facility (ALCF) (US DOE Contracts
No.~DE-AC02-05CH11231 and DE-AC02-06CH11357) and under an INCITE award
(US DOE Office of Advanced
Scientific Computing Research).
\end{acknowledgement}


\providecommand{\APSLONG}{}
\providecommand{\ELSEVIER}{}


\end{document}